\documentclass[preprint2]{aastex}

\def\t0{\theta_{\circ}}

\def\be{\begin{equation}}
\def\en{\end{equation}}

\def\etal{et al.\ }

\def\lsun{L_{\sun}}
\def\msunyr{M_{\sun}\,yr^{-1}}

\def\gccm{\rm \, g\, cm^{-3}} 
\def\mdot{\dot{M}}

\slugcomment{To appear in Ap J Spitzer issue}

\shorttitle{IRAC colors of YSO}
\shortauthors{Allen et al.}

\begin{document}


\title{IRAC Colors of Young Stellar Objects}


\author{Lori E. Allen\altaffilmark{1}, Nuria Calvet\altaffilmark{1}, 
Paola D'Alessio\altaffilmark{2}, Bruno Merin\altaffilmark{3}, 
Lee Hartmann\altaffilmark{1}, S. Thomas Megeath\altaffilmark{1}, 
Robert A. Gutermuth\altaffilmark{4},   
James Muzerolle\altaffilmark{5}, Judith L. Pipher\altaffilmark{4}, 
Philip C. Myers\altaffilmark{1}, 
Giovanni G. Fazio\altaffilmark{1}} 

\email{lallen@cfa.harvard.edu}




\altaffiltext{1}{Harvard-Smithsonian Center for Astrophysics, 
    60 Garden Street, Cambridge, MA 02138}
\altaffiltext{2}{Centro de Radioastronom\'\i a y Astrof\'\i sica,
Ap.P. 72-3 (Xangari), 58089 Morelia, Michoac\'an, M\'exico}
\altaffiltext{3}{LAEFF, VILSPA, Apartado de Correos 50727, 28080 Madrid, Espa\~na}
\altaffiltext{4}{Department of Astronomy, University of Rochester} 
\altaffiltext{5}{Steward Observatory, University of Arizona}


\begin{abstract}
We compare the infrared colors predicted by theoretical models of
protostellar envelopes and protoplanetary disks with initial
observations of young stellar objects made with the Infrared Array
Camera (IRAC) on the Spitzer Space Telescope (Werner et al. 2004, 
Fazio et al. 2004).  Disk and envelope models
characterized by infall and/or accretion rates found in previous
studies can quantitatively account for the range of IRAC colors found
in four young embedded clusters: S~140, S~171, NGC~7129, and Cep~C.  
The IRAC color-color diagram ($[3.6]-[4.5] \ vs. \ [5.8]-[8.0]$) can be
used to help to distinguish between young stars with only disk emission
and protostars with circumstellar envelopes. 
\end{abstract}





\section{Introduction}

It has now been several decades since the first observations of infrared 
excess emission from young stars (Mendoza 1966, 1968). 
The excess emission is well above that expected from reddened stellar 
photospheres and originates from the dusty circumstellar disks and envelopes
surrounding young stars. For these reasons, infrared color-color diagrams have   
proven to be excellent tools for identifying and classifying young stellar objects (YSO).  
In general, young stars are   
found to fall in three regions in the near-IR (JHKL) diagrams. Objects with   
accretion disks (Class II) fall on the classical T Tauri (CTT) locus or along the 
reddened CTT locus, and objects whose emission is dominated by an infalling envelope 
(Class I) fall redward of the reddened CTT locus. Stars having disks with 
large inner holes 
are found in the region corresponding to reddened main sequence stars (Meyer, Calvet 
\& Hillenbrand 1997). 
In an exhaustive study of star formation in the 
Taurus molecular cloud, Kenyon \& Hartmann (1995) combined ground-based 
near-IR photometry with IRAS fluxes to derive spectral indices for 
Class I and II sources and to show that there is a smooth progression in 
IR colors from disk-dominated Class II to envelope-dominated Class I. 
They, along with others (Lada et al. 2000) also showed that the 
K-L color index is a more effective measure of near-IR excess than H-K  
and is better for distinguishing Class I from Class II sources.  

The Infrared Array Camera (IRAC) on the Spitzer Space Telescope has the potential 
to extend our understanding of disk evolution and star formation by detecting 
optically obscured, deeply embedded young stars and protostars, the emission 
from their disks and, at earlier stages, from their infalling envelopes. 
The great advantage of IRAC over ground-based telescopes is its sensitivity in   
the 3-8 $\mu$m bands that contain relatively little contribution from stellar 
photospheres as compared to disks and envelopes. It is important that we 
understand this new color space and how to use it to identify young stars of 
various evolutionary classes.  
This contribution presents a preliminary interpretation of the IRAC color-color diagram, 
using predictions of existing models
for disks and envelopes, and adopting
values for parameters which are well understood  
from star formation studies of 
nearby regions like the Taurus molecular cloud.
These models define
clearly separated regions in the
IRAC color-color diagram.
IRAC observations of
four young clusters (Megeath et al. 2004) are consistent with the model predictions.

\section{Models}
\label{models}

Models in the disk grid were calculated
according to the procedures of D'Alessio et al. (1998, 1999, 2001).
In brief, the disk is assumed to be steadily
accreting at a rate $\mdot$ onto  
a star of age $t$, mass $M$, radius $R$, and effective temperature $T_{eff}$.
The material in the disk consists of gas
and dust, with the standard mass ratio ($\rm M_{dust}/M_{gas} = 10^{-2}$), well
mixed and uniformly distributed. 
The dust mixture is that proposed by Pollack et al. (1994)
and has a size distribution $n(a) da \propto a^{-p} da$
between limiting sizes $a_{min}$ and $a_{max}$.
The disk is heated by viscous dissipation
and by irradiation from the central object,
and viscosity is calculated with the $\alpha$
prescription (Shakura \& Sunyaev 1974).
Models are truncated at the dust destruction
radius $R_{rim}$, where the disk is frontally illuminated
by the central object, because gas inside $R_{rim}$ is 
optically thin (D'Alessio et al. 2004). The truncation radius
is set by the sum of the stellar and accretion
luminosity (Muzerolle et al. 2003), and the dust 
destruction temperature
is set at 1400K, the sublimation temperature
of silicates at characteristic densities
of the inner disk. The wall at $R_{rim}$, which 
has a fixed height of four scale heights,  
emits as a black body at this temperature
(Natta et al. 2001; Muzerolle et al. 2003; D'Alessio et al. 2003).
The equations
of disk structure are solved including these
heating sources to yield the detailed radial-vertical
structure. The emerging spectral energy distribution (SED)
is calculated by ray-by-ray integration
of the transfer equation for 
each line of sight.
Using these procedures, we have constructed an
extensive grid of disk models which 
cover the following range of parameter space:
$T_{eff} =$ 4000K - 10000K for 1 and 10 Myr old stars,
log $\mdot = -9$ to $-6\, (\msunyr)$,
disk radii $R_d$ = 100 and 300 AU,
$p = 2.5, 3.5$,  
$a_{max} = 1 - 10^{5} \micron$, and
inclinations $30^{\circ}$ and $60^{\circ}$,
with fixed $a_{min} = 0.005 \micron$ and $\alpha = 0.01$.
Details of the grid will be published elsewhere
(D'Alessio et al. 2004). Here we present only those models 
for $T_{eff} =$ 4000K, $t = 1$ Myr.  

Models in the Class I grid were calculated
following the procedures of Kenyon, Calvet, \& Hartmann (1993, KCH93)
and Calvet et al. (1994).
The slowly rotating infalling envelope structure of Terebey, Shu, \& Cassen (1984)
has been adopted; in this model, the envelope
is nearly spherically symmetric at radii much
larger than the centrifugal radius $R_c$  
and departs from spherical symmetry at radii $< R_c$,
where material falls onto the disk. The
heating of the matter in the envelope with
density distribution $\rho(r,\theta)$ is 
set by the luminosity of the central object $L$. The temperature
is calculated from the radiative equilibrium condition
using the angle-averaged density, and the inclination
dependent flux is calculated from ray-by-ray integration
of the transfer equation using the angle-dependent density.
The scattering component of the source function
in the flux calculation has been taken as 
the sum of the direct and the diffuse mean intensity,
following Calvet et al. (1994).
Class I models were calculated for
$L=0.1, 1, 10$ and $100 \lsun$,
log $\rho_1=-14, -13.5, -13.35, -13, -12.75$ and $-12.5\, (\gccm)$,
where $\rho_1$ is the density at 1 AU of the equivalent
spherically symmetric envelope (see KCH93), and
$R_c$ = 50 and 300 AU. Model colors shown
have been calculated for an inclination of 60$^\circ$.

Model colors were calculated by convolving the model spectral energy 
distribution with the IRAC filter response functions and by using the 
Vega fluxes in the IRAC bands as photometric zero points. The resulting 
magnitudes are thus referenced to the Vega system (as are the measured 
magnitudes of the embedded cluster sample).
Figure 1 shows IRAC colors from grids
of disk (Class II) and infalling envelope (Class I)
models. The models cover the range of parameters
typical of young stellar objects. 

\section{Discussion}

\subsection{Disk models}

The novel element introduced in the disk models
presented here is the emission of the disk
wall at the dust sublimation radius.
A sharp transition in the disk where dust
sublimates, illuminated directly by the star,  
was presented as the explanation
for the near infrared SEDs of Herbig Ae/Be
stars by Natta et al. (1999). Muzerolle et al. (2003)
found similar excesses in Classical T Tauri stars
and realized that
for these low luminosity objects, the accretion
luminosity $L_{acc}$ emitted by the accretion shock
on the stellar surface must be included
when finding the
dust sublimation radius.

Figure 2 shows the SEDs for a representative
model.
The wall contribution
plays a substantial role in the IRAC
wavelength range and results in
a relatively compact region in the $[5.8]-[8.0] \ vs. \
[3.6]-[4.5]$ color plane, as shown by the triangles in Figure 1.

The spread in the $[3.6]-[4.5]$ color is dominated by accretion rate; sources
with higher $\mdot$  are redder in $[3.6]-[4.5]$. 
This behavior is due to an increase of
both the disk emission and the wall emission
as $\mdot$ increases. As the accretion rate rises, 
there is an increase
in viscous dissipation resulting in
higher disk fluxes. In addition, the increase 
in $L_{acc}$ results in more energy irradiating the
wall; as a result, the dust sublimation radius
moves outward, and the wall emitting area increases
(Muzerolle et al. 2003).
Although not shown here for economy of space, 
we note that the wall emission becomes increasingly   
dominant as the stellar luminosity increases, and
so does the wall emitting area. In fact, for
stellar properties corresponding to Herbig Ae/Be
stars, for which $L_* > L_{acc}$, 
the predicted $[3.6]-[4.5]$ color is approximately constant, $\sim$ 0.4,
equal to that of the black body at T = 1400K, the
assumed dust destruction temperature.

Disk emission is more conspicuous in the $[5.8]-[8.0]$ color,
as shown in Figure 2, and therefore disk properties such as
grain size and $\mdot$
play a larger role in determining the
spread. For instance, the disk contribution
decreases as $a_{max}$ increases because the
disk becomes less flared (D'Alessio et al. 2001).
As a result, the $[5.8]-[8.0]$ colors becomes bluer.
Another source of spread is inclination.
The two inclinations plotted in Figure 1 ($30^{\circ}$ and $60^{\circ}$)
form discernible
loci, with the higher inclination sources redward of the lower ones.
Although a Class II source with an extremely high inclination (i.e. edge-on)
may lie blueward of this region (reflecting the scattered
stellar component), most of the inclination-dependent
behavior of the models is encompassed in the region plotted.

\subsection{Envelope models}

Envelope models (plotted as circles in Figure 1) 
span a larger space in both colors
compared to disk models.
To understand the color behavior in
a schematic way consider Figure 3, which
shows model SEDs. 
An increase in density in the
envelope results in a shift of the $\tau \sim 1 $ surface
further out in the envelope (cf. Appendix
in KCH93; Hartmann 1998). As a result, the overall
SED shifts toward longer wavelengths,  
and the $[3.6]-[4.5]$ colors become rapidly redder
with increasing density. 
In addition,
as the amount of material in the envelope increases, 
absorption increases, and the silicate feature at
$\sim$ 10 $\micron$ becomes deeper.
The $[5.8]-[8.0]$ color becomes first bluer, as the
flux in [8.0], contaminated by the silicate
feature, decreases faster than the flux in the [5.8]
band. As density continues to increase, this situation
reverses, and the $[5.8]-[8.0]$ color becomes redder.

As shown in the bottom panel of Figure 3,
an increase in $R_c$, which implies clearing
of the inner envelope, is equivalent in the
near- to mid-infrared to a decrease in density.
As a result, the $[3.6]-[4.5]$ color becomes
bluer, and so does   
the $[5.8]-[8.0]$ color for high values of $\rho_1$. However,
for low $\rho_1$, the silicate feature turns
into emission, making the
$[5.8]-[8.0]$ color redder (Figure 1).

The color behavior as a function of $\rho_1$ and $R_c$ does not depend 
strongly on source luminosity. However, 
as shown in Figure 2
of KCH93, the strength of the silicate
absorption feature decreases as luminosity increases.
At high values of $\rho_1$,
the $[5.8]-[8.0]$ colors are less
affected by the silicate absorption and do not
get as blue as $L$ increases. At low densities,
the extinguished spectrum of the central object begins
to contribute along with envelope emission in the
IRAC bands, which tends to make the color bluer; a competing
effect is the increase in strength
of the silicate emission, which makes $[5.8]-[8.0]$ redder.
For $L = 0.1 \lsun$, the second effect is negligible
so colors are very blue (Figure 1).

Other effects not included in this preliminary modeling effort,
such as degree of flattening of the envelope or the presence of outflow
cavities, may change the colors as well, especially
at low densities.  For example,   
Whitney \etal (2003) predicted IRAC colors for a range of
protostellar (Class 0,I) and T Tauri objects (Class II).
Their calculations for Class II objects span a much smaller range in
the color-color diagram because they considered a much smaller range of
parameter space than we have here, and they did not include the
effects of a hot inner disk wall.  The objects Whitney et al. label as
Class I have predicted colors that are considerably bluer than we infer
here and 
the region populated by Class 0 sources in the
Whitney \etal calculations overlaps with the region of Taurus Class I
sources (Hartmann et al. 2004).  The discrepancy appears to be the
result of the particular outflow cavities adopted by Whitney \etal,
which greatly reduce the amount of dust close to the central source and
thus strongly reduce the extinction and dust thermal emission in the
3-8 $\mu$m region (see Osorio \etal 2003 for a related discussion).

\subsection{Comparison with observed colors}

Figure 4 shows measured IRAC colors of four young clusters (Megeath et al. 2004; 
Gutermuth et al. 2004) and for 
comparison the loci of the models presented in Figure 1. 
The data seem to cluster into three main regions: a clump around 0,0 that 
contains mostly background/foreground stars and Class III sources with no 
intrinsic infrared excess, a clump that occupies the   
Class II region (within the large blue box), 
and a group that runs along the Class I locus  
with $L \ \ge \ 1 \lsun$. 
Sources which lie between the $[5.8]-[8.0]$ colors of stellar photospheres 
and Class III stars ($\sim$ 0) and Class II sources ($\ge$ 0.4) 
can be understood if the height of the wall at $R_{rim}$ was less than 
assumed here, due to the effects of dust grain growth and settling. 
In addition, there are a few sources which do not lie inside the Class II 
locus but {\it do} lie along the extinction vector and could therefore be 
reddened Class IIs. The most extreme of these, near $[5.8]-[8.0] = 0,   
[3.6]-[4.5] = 1.5$, are coincident with filaments of high opacity (dark even 
at 8 $\mu$m), consistent with the large $A_v$ implied by this scenario. 
Alternatively they could be extremely low-luminosity Class I sources. 

The common definition of Class I and II sources is that
they have positive or negative spectral indices, respectively, in
$\lambda F_{\lambda}$; thus the approximate boundary between Class I
and II should lie approximately at $[3.6]-[4.5] \sim 0.7$ and
$[5.8]-[8.0] \sim 1$. The model predictions presented here are 
in agreement with this boundary. 
As discussed, features in the
SED, notably the silicate feature at $\sim$ 10 $\micron$,
have a large effect on the $[5.8]-[8.0]$ color, producing 
some exceptions to this rule. 
In addition, the $[3.6]-[4.5]$ colors of low density envelopes 
are very sensitive to the nature of the central objects contained within. 
For example, low density envelopes around stars
with disks would have $[3.6]-[4.5]$ colors near the Class II locus.

\section{Summary}

The IRAC color-color diagram ($[5.8]-[8.0]\ vs.\ [3.6]-[4.5]$) 
is a useful tool for identifying young stars having infrared excess 
emission.  While there is some overlap between low density, low luminosity 
Class I sources and Class II sources, the classes are otherwise well segregated. 
The distribution of measured IRAC colors of sources in four young clusters 
is consistent with the distribution of model colors. Confirmation of this 
color-color diagram as a powerful diagnostic tool awaits a more detailed 
comparison to the observed colors of well characterized individual sources. 
 
\acknowledgments
This work is based on observations made with the Spitzer Space Telescope, which is
operated by the Jet Propulsion Laboratory, California Institute of Technology under
NASA contract 1407. Support for this work was provided by NASA through Contract Number
1256790 issued by JPL/Caltech.
This work was supported in part by grant AR-09524.01-A from the
Space Telescope Science Institute, and by NASA Origins of Solar Systems
grant NAG5-9670.


Facilities: \facility{Spitzer Space Telescope}

\clearpage



\onecolumn
\begin{figure}
\plotone{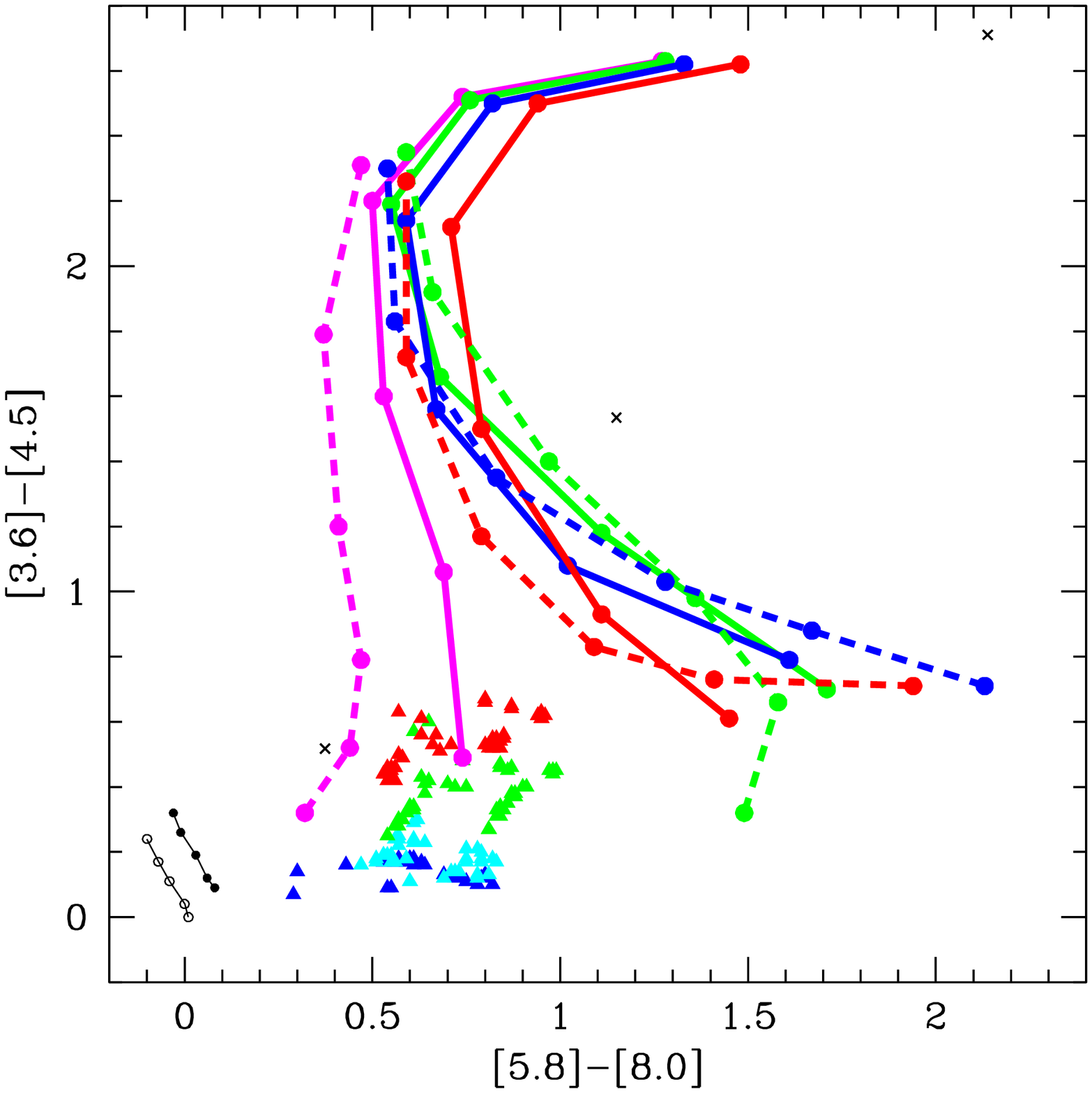}
\caption{IRAC colors for disk models (triangles) and envelope models (circles). 
{\bf Disk models} 
assume $T_{eff} = 4000 K, t = 1 Myr$. Disk accretion rates are color coded: 
dark blue, light blue, green 
and red represent log $\mdot = -9, -8, -7, -6\, (\msunyr)$, respectively.  
Models are plotted for two values of $p$ (2.5, 3.5) and 
six values of $a_{max}$ (1$\mu$m, 10$\mu$m, 100$\mu$m, 
1mm, 1cm and 10cm).  
The two inclinations ($30^{\circ}$ and $60^{\circ}$)   
form discernible loci centered at $[5.8]-[8.0]=0.6$ and 0.9. \
{\bf Envelope models} are shown for a range of central source 
luminosities, color coded as magenta, 
green, blue, red = 0.1, 1, 10 and 100 $\lsun$, respectively. 
Envelope densities are shown for 
log $\rho_1 = -14, -13.5, -13.35, -13, -12.75$, and $-12.5\, (\gccm)$, 
increasing from bottom to top. 
Models are plotted for two values of $R_c$; 50 AU (solid line), 
and 300 AU (dashed line), 
and one inclination ($60^{\circ}$). 
For comparison, blackbody colors (T=300K, 500K, 1200K) are plotted (as x's). 
Two blackbodies with extinctions of $A_v = 0,5,10,15,20\,$mag 
are shown; open circles correspond to a blackbody of T=10,000K, 
filled circles to T=4,000K. 
\label{fig1}} 
\end{figure}

\clearpage

\begin{figure} 
\plotone{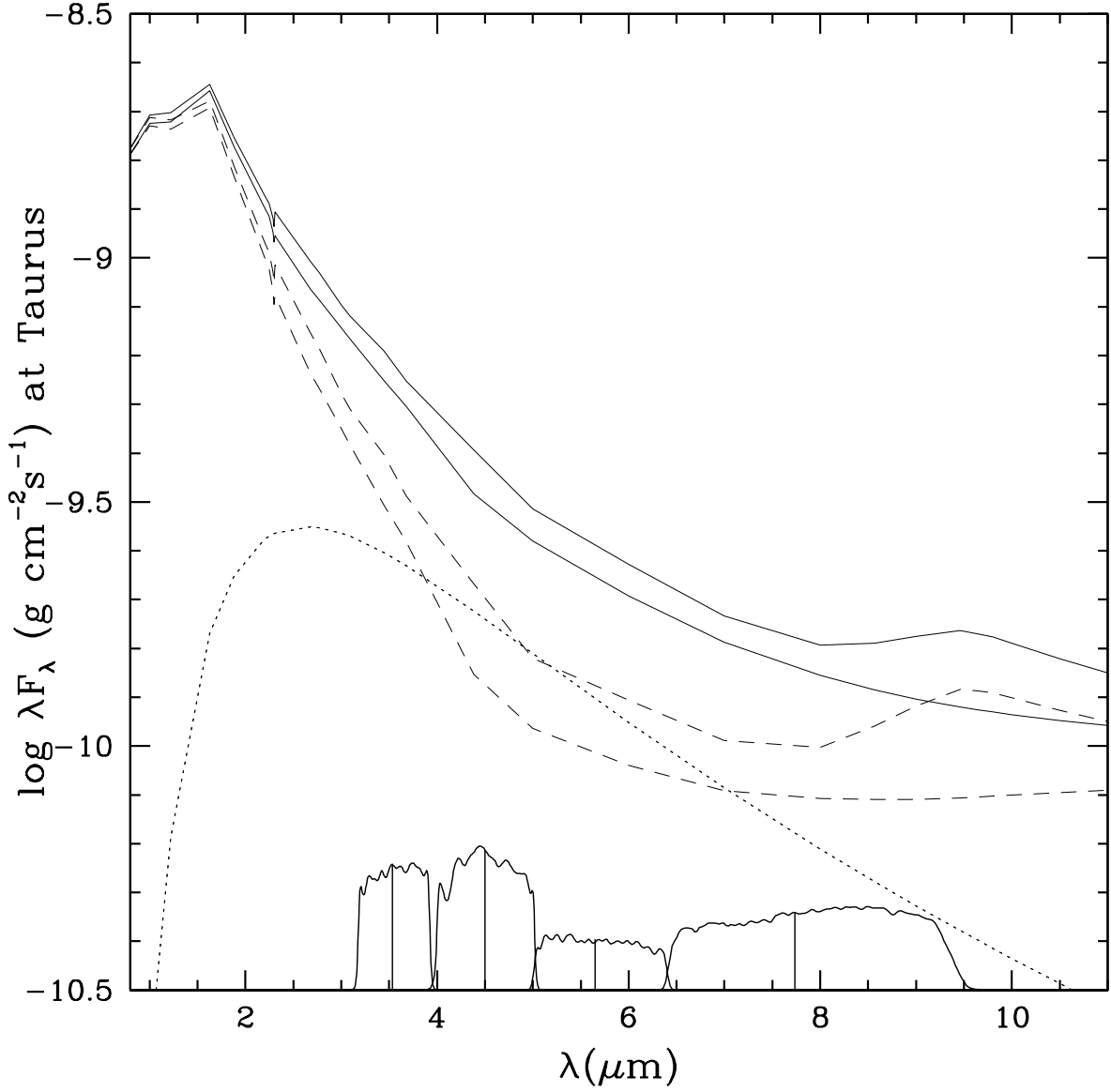}
\caption{Representative SEDs for a disk model with inclination $= 60 ^{\circ}$, $\mdot = 10^{-8} \msunyr$ 
and two values of 
$a_{max}$, $0.1 \micron$ (upper) and 1mm (lower). The solid lines show the total 
emission (star $+$ disk $+$ wall), the dashed lines show the emission from the 
star$+$disk, and the dotted line shows the emission from the wall, modeled as a 1400 K 
blackbody.  Spectral response curves for the IRAC bands (normalized to an 
arbitrary value) are shown at bottom, and the band effective wavelengths are 
indicated (vertical lines). The wall contributes 
significantly to the total emission, especially in the IRAC bands.
\label{fig2}}
\end{figure}

\clearpage

\begin{figure} 
\plotone{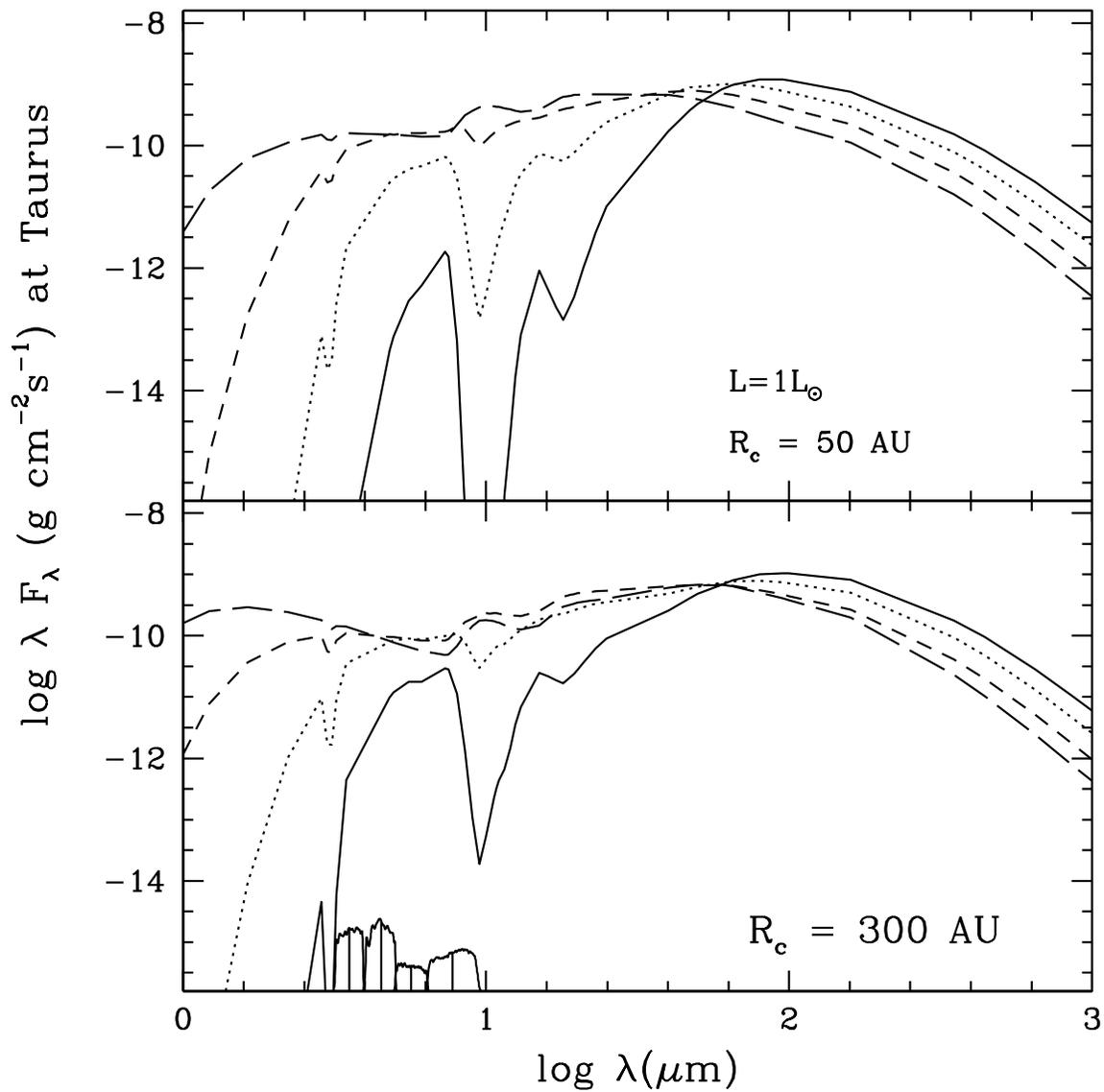}
\caption{Representative SEDs for envelope models with
$L = 1 \lsun$  and $R_c  = 50$ AU (upper panel)
and $R_c  = 300$ AU (lower panel), and for densities
$\log \ \rho_1 = -14, -13.5, -13$, and $-12.5\, (\gccm)$,
increasing from top to bottom. As in the previous figure, 
IRAC bandpasses are indicated.
\label{fig3}}
\end{figure}

\clearpage

\begin{figure}
\plotone{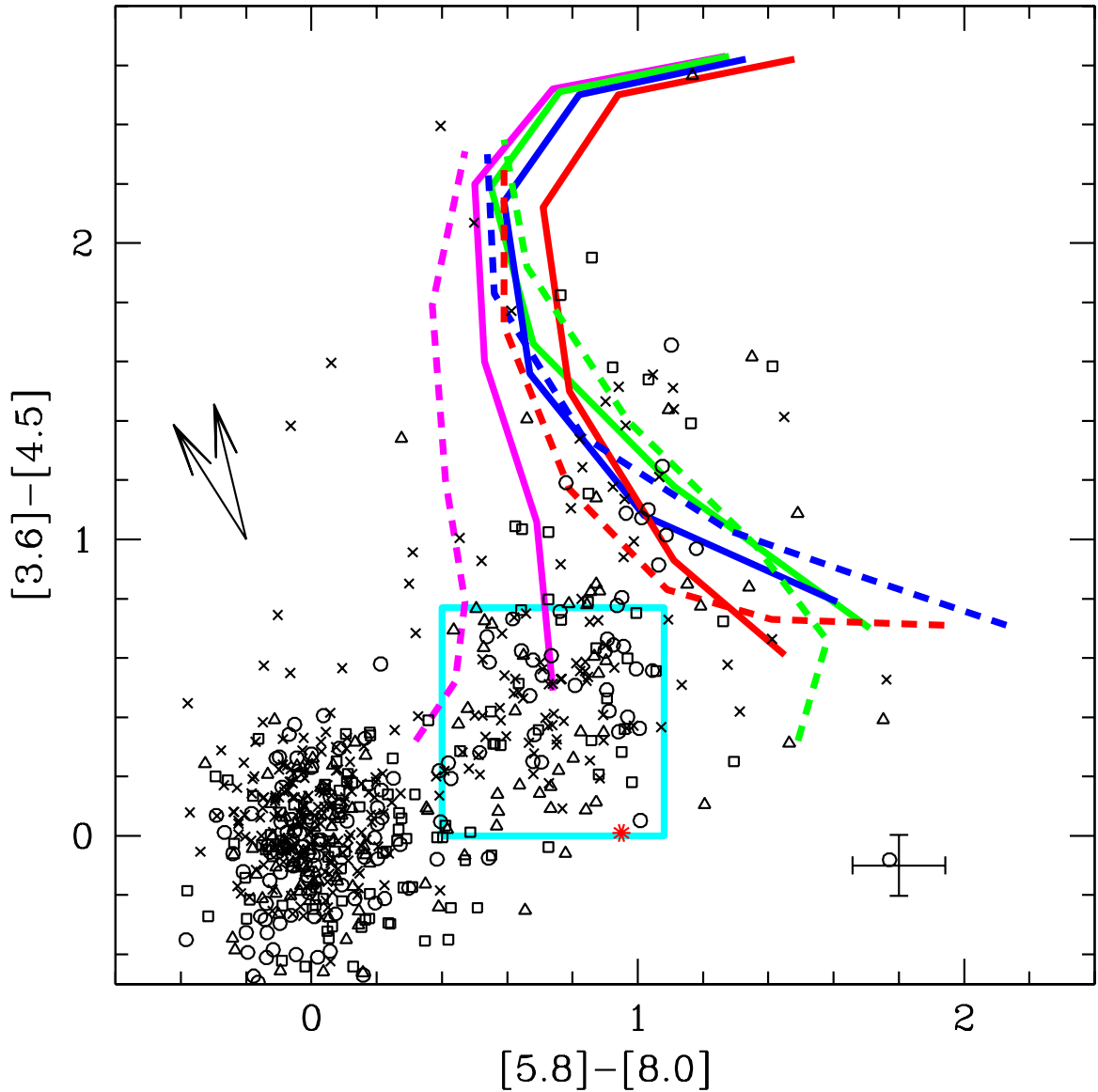}
\caption{Model colors from Figure 1, and measured IRAC colors for 
S~140 (squares), Cep~C (x's), S~171 (circles) and NGC~7129 (triangles)  
from the IRAC GTO embedded clusters survey (Megeath et al. 2004). 
Representative error bars, including average photometric uncertainties and
an estimated 10\% uncertainty in the absolute flux calibration, are shown.
The light blue square delineates the approximate domain of 
Class II sources, and the colored lines show the Class I models as in Figure 1. 
The red asterisk marks the IRAC colors of TW Hya (Hartmann et al. 2004), a 10 
Myr old star with a disk (Muzerolle et al. 2000). 
Extinction vectors are shown for $A_v = 30\,$mag, using the two extremes of   
the six vectors calculated by Megeath et al (2004).  The vector on the left is for 
a flat spectrum source and Draine \& Lee (1984) extinction law, while the vector to 
the right is for a Vega spectrum and the extinction law of Mathis (1990). 
\label{fig4}}
\end{figure}

\end{document}